# Temperature excitation of the Co$^{3+}$ triplet spin-state in LaCoO$_3$ determined by polarized neutron diffraction

V P Plakhty$^{1,*}$, P J Brown$^2$, B Grenier$^3$, S V Shiryaev$^4$, S N Barilo$^4$, S V Gavrilov$^1$ and E Ressouche$^3$

$^1$Petersburg Nuclear Physics Institute, 188300, Gatchina, St. Petersburg, Russia
$^2$Institut Laue-Langevin, 38042 Grenoble Cedex 9, France
$^3$CEA-Grenoble, DFMC/SPSMS/MDN, 38042 Grenoble Cedex 9, France
$^4$Institute of Solid State & Semiconductor Physics, 220072, Minsk, Belarus

**Abstract**
The magnetic moment induced by an external magnetic field of 5.8 T on the Co$^{3+}$ ions in LaCoO$_3$ has been measured by polarized neutron diffraction. Measurements were made at 100 K, near to the spin-state transition, at 1.5 K and at 285 K. The data give evidence for a transition from the ground singlet state ($S = 0$) to the excited triplet intermediate spin state ($S = 1$). The energy of $\Delta_\mathrm{H} \approx 1000$ K estimated for the high-spin (S =2) state, is too high to contribute significantly even at $T = 285$ K. There is evidence for a small negative covalent spin density associated with the oxygen ligands amounting to about one fifth of the aligned Co$^{3+}$ moment at all three temperatures.

## 1. Introduction

Competition between intra-atomic exchange and the crystal field results in three possible states for Co$^{3+}$ ions in solids: the low-spin state (LS, $t_{2g}^6 e_g^0$, $S = 0$), the intermediate-spin state (IS, $t_{2g}^5 e_g^1$, $S = 1$) and the high-spin state (HS, $t_{2g}^4 e_g^2$, $S = 2$) [1]. The small energy differences between these states, together with the presence of a semiconductor-to-metal transition, leads to the quite unusual magnetic properties of LaCoO$_3$. Although this material has been extensively studied during the last 40 years, the nature of the lowest energy excited spin state remains unclear. The magnetic susceptibility of LaCoO$_3$ is negligible at temperatures $T < 50$ K, goes through a maximum around 100 K then decreases following the Curie-Weiss law to a plateau in the range 400 < T < 600 K characteristic of a semiconductor-to-metal transition [2]. The effective moments of the $Co^{3+}$ ions calculated from the susceptibility data [2,3] are $M \approx 2\,\mu_B$ at $T \approx 100$ K and $M \approx 3.5\,\mu_B$ at $T \approx 900$ K. No long-range magnetic order has been found in all the temperature range where $M \neq 0$ [4,5], although short-range ferromagnetic correlations have been observed in neutron polarization analysis experiments [6]. The susceptibility maximum at 100 K was originally thought to arise from the LS – HS transition [7] and the insulating properties below 400 – 600 K were attributed to NaCl-type ordering of LS and HS $Co^{3+}$ ions [8]. However this latter hypothesis is not supported by neutron diffraction studies [9].

$^*$ Author to whom any correspondence should be addressed: plakhty@pnpi.spb.ru

A completely different approach [10] has been made based on the observation that oxides with formally high oxidation states may be negative charge-transfer systems. For $Co^{3+}$ a relatively stable hole state on the ligand ions favours the IS state [11,12]. The IS state has the configuration $t_{2g}^5 e_g^1$, in which the double degeneracy of the $e_g$ orbitals is lifted by the Jahn-Teller effect. In the rhombohedral lattice containing two Co ions per unit cell, "antiferro" orbital ordering of the half-filled orbitals, for instance $y^2-z^2$, $yz$ and $x^2-y^2$, $xy$, minimizes the Coulomb interaction between $e_g$ and $t_{2g}$ electrons [10]. Such ordering could account for the lack of metallic conduction in the range $100 < T < 500$ K. According to [10], the broad semiconductor-to-metal transition around 500 K is associated with gradual orbital disordering of IS state ions, with the transition to the HS state taking place at yet higher temperature. This idea that the anomalous magnetic and thermal behaviour is due to a LS – IS transition accompanied by orbital ordering has motivated a number of recent experiments. Magnetic susceptibility and thermal expansion [13], susceptibility, photoemission spectroscopy and x-ray absorption spectroscopy [14], electron spin resonance data [15] and structural properties [16,17] are all in reasonable agreement with the model [10]. The agreement can be improved by introducing further parameters: for instance, a three-level model with a varying population of LS, IS and HS-states has been explored in [16] to describe both structural and magnetic data. The phonon modes in Ref. [18] were interpreted in terms of a dynamical Jahn-Teller effect in the thermally excited IS-state. A cooperative Jahn-Teller distortion associated with $e_g$ orbital ordering [10] was later demonstrated by high-resolution x-ray diffraction [19]; the space group symmetry was shown fall from $R\bar{3}c$ to its monoclinic subgroup $I2/a$ on heating through $T \approx 100$ K. However discussion is still going on about whether the first excited state is IS or HS.

It is clear that an important point to understand is the lack of long-range magnetic order above the spin-state transition independently on the excited state. The experimental data on both the magnetic susceptibility and specific heat of $LaCoO_3$ can be accounted for if the spin-state excitation is assumed to reduce the exchange interaction [20]. This effect is supposed to originate from a difference in the covalence between Co-O bonds formed by low-spin and excited spin $Co^{3+}$ ions. Six models involving mixtures of IS and HS states with different degeneracies and $g$-factors were tested in [20]. Whilst all models could reproduce the magnetic susceptibility up to 300 K, the specific heat data up to 100 K could be accounted for by an exited state corresponding to only one of the IS + HS mixtures [20].

The objective of our experiment was to measure the moment on the $Co^{3+}$ ion and on the oxygen ligands below and above the spin-state transition, to show whether or not the absence of magnetic ordering is due to a negative cooperative effect, i.e., weaker superexchange in the covalent bond for the excited spin states in comparison with the ground LS state. The measurements of the $Co^{3+}$ moment at $T = 100$ K and $T = 285$ K should also allow distinction between the two scenarios for the spin-state transition, LS $\rightarrow$ HS and LS $\rightarrow$ IS.

## 2. Polarized neutron diffraction and its application to LaCoO$_3$

The intensity of unpolarized neutrons scattered by a Bragg reflection with scattering vector **k** from a crystal with ordered magnetic moments can be expressed as

$$I(\mathbf{k}) \propto |F_N|^2 + |\mathbf{M}_\perp|^2, \qquad (1)$$

where the nuclear, $F_N$, and magnetic, $\mathbf{F}_M$, structure factors are given in terms of the coordinates $\mathbf{r}_i$ of atoms in the unit cell, their nuclear scattering lengths $b_i$, their magnetic moments $\mathbf{m}_i$ and magnetic form-factors $f_i(\mathbf{k})$ as

$$F_N = \sum_i b_i \exp(i\mathbf{k}\cdot\mathbf{r}_i), \qquad (2)$$

$$\mathbf{M}_\perp = \mathbf{k} \times \mathbf{F}_M \times \mathbf{k} \qquad (3)$$

with

$$\mathbf{F}_M = \sum_i \mathbf{m}_i f(\mathbf{k}) \exp(i\mathbf{k}\cdot\mathbf{r}_i). \qquad (4)$$

For neutrons with initial polarization **P** parallel to the applied field

$$I^+(\mathbf{k}) \propto |F_N|^2 + 2\Re(F_N^*(\mathbf{P}\mathbf{M}_\perp)) + |\mathbf{M}_\perp|^2. \qquad (5)$$

When the structure is centro-symmetric and the magnetic moments are aligned parallel to **P** by a magnetic field, $F_N$ and $\mathbf{F}_M$ are real. If the ratio between the magnetic and nuclear structure factors is $\gamma$, the unpolarized intensity $I(\mathbf{k}) \propto 1+\gamma^2$ and the ratio $R$ between the intensities of neutrons with polarization parallel and antiparallel to the applied field is

$$R = \frac{I^+}{I^-} = \frac{|F_N|^2 + 2F_N\mathbf{P}\cdot\mathbf{M}_\perp + |\mathbf{M}_\perp|^2}{|F_N|^2 - 2F_N\mathbf{P}\cdot\mathbf{M}_\perp + |\mathbf{M}_\perp|^2} = \frac{1+2P\gamma+\gamma^2}{1-2P\gamma+\gamma^2}. \qquad (6)$$

When $\gamma$ is small the contribution of magnetism to the unpolarized intensity is proportional to $\gamma^2$, whereas its contribution to $R$ is proportional to $4P\gamma$. Flipping ratio measurements therefore make it possible to determine very small magnetic moments such as those induced by an external field in a paramagnetic crystal.

According to a recent neutron diffraction study [17], the structure of LaCoO$_3$ in the temperature range $5 \leq T\,(\mathrm{K}) \leq 1000$ is that of rhombohedrally distorted perovskite with space group $R\bar{3}c$. The rhombohedral unit cell, with edges $\mathbf{a}_R$, $\mathbf{b}_R$ and $\mathbf{c}_R$, shown in figure 1 contains two perovskite cells with edges $\mathbf{a}_C$, $\mathbf{b}_C$, $\mathbf{c}_C$, orientated so that $\mathbf{a}_R = \mathbf{a}_C + \mathbf{b}_C$, $\mathbf{b}_R = \mathbf{b}_C + \mathbf{c}_C$ and $\mathbf{c}_R = \mathbf{c}_C + \mathbf{a}_C$. In the hexagonal description of the rhombohedral cell, which is that used in the rest of this paper, the cell edges are $\mathbf{a}_H = \mathbf{a}_C - \mathbf{c}_C$, $\mathbf{b}_H = -\mathbf{a}_C + \mathbf{b}_C$, $\mathbf{c}_H = 2\mathbf{a}_H + 2\mathbf{b}_H + 2\mathbf{c}_H$.

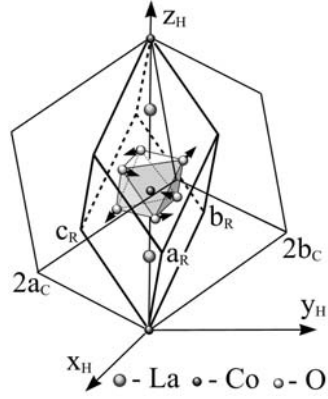

**Figure 1.** Unit cell of LaCoO$_3$. The explanations are given in the text.

The doubling of the perovskite cell is exclusively due to antiparallel displacements of the oxygen atoms as shown by the arrows in figure 1. This means that nuclear structure factors of reflections with $l = 2n+1$ are due entirely to the oxygen ions, whereas all the constituent atoms contribute to those with even $l$. The magnetic structure factors of the even $l$ reflections depend mainly on the magnetic moment on the Co ion but will also contain contributions from any moment on the oxygen ligands; those of the odd-$l$ reflections arise only from the perturbation of the atomic magnetization due to the rhombohedral distortion.

The rhombohedral distortion is due to compression along one of the four body diagonals of the cubic perovskite cell. Any one of these diagonals may be chosen resulting in 4 possible twins; the number twins being equal to the ratio between the orders of the symmetric group $Pm3m$ – 48 and its sub-group $R\bar{3}c$ – 12. In the rhombohedral structure each reflection of the perovskite cell splits into four parts, one from each twin. The matrices relating the indices of these reflections in hexagonal indices are:

$$T_1 = \begin{pmatrix} 1 & 0 & 0 \\ 0 & 1 & 0 \\ 0 & 0 & 1 \end{pmatrix},\ T_2 = \begin{pmatrix} -\tfrac{1}{3} & -\tfrac{2}{3} & \tfrac{1}{3} \\ 1 & 1 & 0 \\ -\tfrac{4}{3} & \tfrac{4}{3} & \tfrac{1}{3} \end{pmatrix},\ T_3 = \begin{pmatrix} -1 & 0 & 0 \\ \tfrac{2}{3} & \tfrac{1}{3} & \tfrac{1}{3} \\ \tfrac{4}{3} & \tfrac{8}{3} & -\tfrac{1}{3} \end{pmatrix},\ T_4 = \begin{pmatrix} \tfrac{1}{3} & \tfrac{2}{3} & -\tfrac{1}{3} \\ -\tfrac{1}{3} & \tfrac{1}{3} & \tfrac{1}{3} \\ \tfrac{8}{3} & \tfrac{4}{3} & \tfrac{1}{3} \end{pmatrix}. \quad (7)$$

Since the distortion is small ($\alpha = 90.98^\circ$ at room temperature) these 4 reflections cannot be resolved by neutron diffractometers with conventional resolution and the twinning must be taken into account in the data analysis. In particular, the twinning will invalidate the trigonal symmetry if the twin populations are not equal.

## 3. Crystal growth, and characterization

Single crystals of $LaCoO_3$ were grown using an anodic electro-deposition technique [22] modified to use seeded flux melt growth based on a solvent consisting of a $Cs_2MoO_4 - MoO_3$ mixture in the ratio 2.2 : 1 [23]. The material from which the crystals were to be grown was placed in a 100 cm$^3$ platinum crucible and an appropriate amount of the solvent added; a seed crystal served as the anode and the crucible itself as the cathode of the two-electrode electrochemical cell. The crystals were grown at a temperature of about 950 – 1000$^o$C under a current density of 0.5 – 0.7 mA/cm$^2$. A crystal of about 400 mg ($4 \times 4 \times 4$ mm$^3$) was chosen for the neutron diffraction experiment. The susceptibility of the crystal was found to have the same Curie-like contribution at low temperature as has been reported in [18,20]. This is probably due to slight oxygen nonstoichiometry, but the contribution of this component to the magnetization at the temperatures of the experiment 100 K and 285 K, 6% and 4%, respectively, are negligible in comparison with the experimental precision of the magnetic moments. A preliminary experiment to test the quality of the crystal and to verify the structural parameters was carried out at ambient temperature on the four-circle diffractometer D15 at ILL with wave-length $\lambda = 1.17$ Å and using a small 32×32 pixel multidetector. The reflections were rather wide and irregularly shaped, which limited the precision of integration; their full widths at half height varied from 2 to 5 degrees of arc. The lattice parameters obtained at $T = 100$ K from refinement of the UB matrix $a_H = b_H = 5.4639(9)$ Å, $c_H = 13.372(4)$ Å are a little bigger than those from the powder diffraction experiment [17], but this is may be due to the calibration of the wavelength. The integrated intensities were used in a least squares refinement to obtain the temperature factors and the coordinate of the oxygen atom. The measured intensities were assumed to contain contributions from all four rhombohedral twins, and the fractions of each twin component present were also refined. The value $x_O = 0.5523(5)$ was obtained for the oxygen position parameter in agreement with [17]. A small imbalance in the domain populations corresponding to fractions 0.230(8), 0.241(9), 0.286(9), 0.244(9) of the twins $T_1$ to $T_4$ of Eq. (7) was present. No significant extinction was found.

## 4. Polarized neutron measurements

The flipping ratio measurements were made on the D23 polarized neutron diffractometer installed at the supermirror-coated thermal neutron guide H25 of the ILL high-flux reactor. D23 uses normal beam geometry with a detector which can be inclined to the horizontal plane. The crystal was mounted with its hexagonal axis [001] vertical and magnetized by a cryomagnet producing vertical field of 5.8 T. Polarized neutrons of wave-length $\lambda = 1.246$ Å were obtained by reflection from the (111) plane of a Heusler alloy monochromator. The flipping ratios of 61, 130 and 260 different Bragg reflections were measured at 1.5 K, 100 K and 285 K, respectively; of these 26 were independent under the rhombohedral symmetry, 9 with $l = 2n+1$ and 17 with $l = 2n$. Table 1 contains the data averaged over rhombohedrally equivalent reflections. In the final analysis these reflections were not averaged together in order to take into account the inequality found in the twin populations.

**Table 1.** Values of the flipping ratios $R$ measured at temperatures of 1.5, 100 and 285 K, averaged over rhombohedrally equivalent reflections.

| h | k | l  | $\sin\theta/\lambda$ | $R$(1.5 K) | $R$(100 K) | $R$(285 K) |
|---|---|----|--------|------------|------------|------------|
| 1 | 1 | 0  | 0.1498 | 1.662(11)  | 1.106(4)   | 1.049(5)   |
| 2 | 1 | −2 | 0.1982 | 0.92(6)    | 0.92(2)    | 0.95(3)    |
| 2 | 0 | 2  | 0.2247 | 0.911(2)   | 0.9799(14) | 0.993(2)   |
| 2 | 2 | 0  | 0.2995 | 1.071(4)   | 1.012(3)   | 1.006(2)   |
| 3 | 0 | 0  | 0.3177 | 1.236(9)   | 1.045(3)   | 1.017(3)   |
| 3 | 2 | −2 | 0.3265 | 0.890(14)  | 0.974(5)   | 0.996(11)  |
| 3 | 1 | 2  | 0.3745 | 1.03(5)    | 0.993(13)  | 1.00(2)    |
| 4 | 0 | −2 | 0.4302 | 0.934(5)   | 0.985(3)   | 0.993(3)   |
| 4 | 1 | 0  | 0.4366 | 1.087(11)  | 1.024(4)   | 1.009(3)   |
| 3 | 3 | 0  | 0.4493 | 1.08(2)    | 1.007(5)   | 1.009(6)   |
| 4 | 3 | −2 | 0.4677 | 0.97(2)    | 0.996(5)   | 1.002(11)  |
| 5 | 1 | −2 | 0.5242 | 1.04(3)    | 1.006(8)   | 1.02(2)    |
| 4 | 2 | 2  | 0.5242 | 0.946(9)   | 0.993(3)   | 0.999(4)   |
| 5 | 0 | 2  | 0.5348 | 0.917(14)  | 0.983(8)   | 0.991(6)   |
| 5 | 2 | 0  | 0.5703 | 1.009(9)   | 1.002(2)   | 1.000(3)   |
| 6 | 0 | 0  | 0.6354 | 1.008(9)   | 1.005(3)   | 0.998(5)   |
| 6 | 1 | 2  | 0.6656 | 1.01(2)    | 0.999(9)   | 1.005(12)  |
|   |   |    |        |            |            |            |
| 2 | 1 | 1  | 0.2621 | 0.994(9)   | 0.996(4)   | 1.007(5)   |
| 1 | 1 | 3  | 0.2622 | 1.008(7)   | 0.998(4)   | 0.997(4)   |
| 3 | 1 | −1 | 0.3199 | 1.000(5)   | 1.003(3)   | 1.001(4)   |
| 3 | 2 | 1  | 0.4119 | 1.00(6)    | 1.006(15)  | 1.00(2)    |
| 2 | 2 | 3  | 0.4119 | 1.009(8)   | 0.999(4)   | 0.997(4)   |
| 4 | 1 | −3 | 0.4119 | 0.999(7)   | 0.995(4)   | 1.005(7)   |
| 4 | 2 | −1 | 0.4509 | 0.998(7)   | 1.004(3)   | 1.001(3)   |
| 4 | 1 | 3  | 0.4868 | 1.004(9)   | 1.000(5)   | 0.998(6)   |
| 3 | 3 | 3  | 0.5617 | 0.994(10)  | 1.000(3)   | 1.004(5)   |

The flipping ratio of each reflection measured was used independently in a least squares refinement of the parameters of various possible models for the atomic magnetization in LaCoO$_3$. Four models of increasing complexity were tried. In the simplest the magnetization is modeled by the spherically symmetric distribution corresponding to a Co$^{3+}$ ion [24] centred at each Co position; the only parameter fitted is the Co$^{3+}$ moment. For the IS state of Co$^{3+}$ the unpaired electron is in an $e_g$ state and there is a hole in the $t_{2g}$ manifold, so the corresponding magnetization should have $e_g$ symmetry with respect to the cubic perovskite cell. In the second model the asphericity appropriate to such $e_g$ symmetry is applied to the Co$^{3+}$ magnetization at each site; again the only parameter fitted is the Co$^{3+}$ moment. The third and fourth models are extensions of the first two in which a small moment with a distribution corresponding to that of the 2$p$ electrons of the O$^{2-}$ ion [24] resides on the oxygen sites. In these latter models two parameters, the Co$^{3+}$ and O$^{2-}$ moments, have been refined. The least squares program calculates the flipping ratio for the model and its differentials with respect to the parameters by accumulating properly weighted contributions to the numerator and denominator of equation (6) calculated from all four twin components. The results obtained with the four models at 1.5, 100 and 285 K are given in table 2. It can be seen that there is not much difference in the goodness of fit at the same temperature between the four models, although in all cases the best fit is obtained with model 3 in which the $Co^{3+}$ moment is spherically symmetric and there is a small negative moment associated with oxygen. There is no evidence for the asymmetry in the $Co^{3+}$ magnetization which would be associated with the IS, $t_{2g}^5 e_g^1$ and HS, $t_{2g}^4 e_g^2$ states although the data are not very sensitive to this asymmetry. The much larger $\chi^2$ for the 1.5 K data is mainly due to the better statistical precision obtained from the larger signal at this temperature.

**Table 2.** Results obtained from least squares refinements of the flipping ratio for four models for the magnetization distribution in LaCoO$_3$

| $T$, K | Parameter | Model 1[1] | Model 2[2] | Model 3[3] | Model 4[4] |
|---|---|---|---|---|---|
| 1.5 | $m(\mathrm{Co}^{3+})$, $\mu_B$ | 0.52(6) | 0.53(6) | 0.41(12) | 0.42(12) |
|  | $m(\mathrm{O}^{2-})$, $\mu_B$ | – | – | −0.11(5) | −0.11(5) |
|  | $\chi^2$ | 56.5 | 56.6 | 43.5 | 43.7 |
| 100 | $m(\mathrm{Co}^{3+})$, $\mu_B$ | 0.129(8) | 0.131(8) | 0.10(2) | 0.11(2) |
|  | $m(\mathrm{O}^{2-})$, $\mu_B$ | – | – | −0.021(7) | −0.021(7) |
|  | $\chi^2$ | 3.81 | 3.83 | 3.19 | 3.23 |
| 285 | $m(\mathrm{Co}^{3+})$, $\mu_B$ | 0.056(4) | 0.057(4) | 0.052(9) | 0.053(9) |
|  | $m(\mathrm{O}^{2-})$, $\mu_B$ | – | – | −0.006(4) | −0.005(4) |
|  | $\chi^2$ | 1.88 | 1.90 | 1.84 | 1.86 |

[1] Spherically symmetric Co$^{3+}$.
[2] Co$^{3+}$ with $e_g$ symmetry.
[3] Spherically symmetric Co$^{3+}$ and spherical O$^{2-}$ 2p magnetization.
[4] Co$^{3+}$ with $e_g$ symmetry and spherical O$^{2-}$ 2p magnetization.

## 4. Results and discussion

The magnetization $m(T)$ of the $Co^{3+}$ ion corrected for the Curie-like contamination measured at 1.5 K gives:

$$m(100\ K) = 0.09(1)\ \mu_B, \quad m(285\ K) = 0.050(9)\ \mu_B. \quad (8)$$

These moment values have been used to calculate the gaps $\Delta_I$ and $\Delta_H$, from the ground LS-state to the IS and HS-states, respectively. These gaps are shown in figure 3, together with the level splitting in an external magnetic field $H$ in multiples of $\delta = g\mu_B H$. The gap obtained from $m(100\ K)$, which is near to the spin-state transition, should be close to 100 K and this criterium can be used to choose which of the two proposed spin states is excited at $T \approx 100$ K.

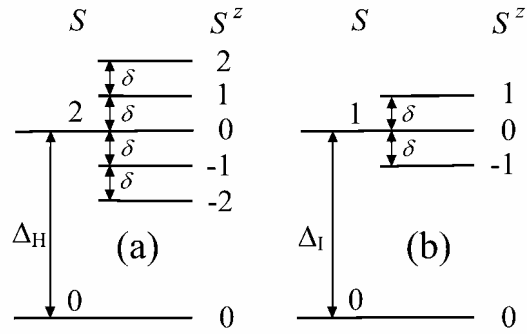

**Figure 3.** The levels splitting for HS state (a) and IS state (b).

One can calculate $m(T)$ as

$$m(T) = -g\mu_B S \nu e^{-\Delta/T} \left\{ \sum_i S_i^z [\exp(\delta_i/T)] \right\} / \left\{ 1 + \nu e^{-\Delta/T} \sum_i \exp(\delta_i/T) \right\}, \quad (9)$$

where $S_i^z$ is the projection of $S$ on the external magnetic field **H**, $\delta_i = g\mu_B S_i^z H$. We use the spin-only value $g = 2$, the sum is over $S(S+1)$ levels $i$, and $\nu$ is degeneracy of the magnetic state. In the model studied in Ref. [10] the IS-state has ordered $e_g$ orbitals $(x^2-z^2)^\uparrow$ and $(y^2-z^2)^\uparrow$ (see figure 6 in Ref. [10]) with the $t_{2g}$ hole in the $(xz)^\downarrow$ and $(yz)^\downarrow$ orbitals; $\nu = 3$ (triply degenerate $t_{2g}$ orbital). In the HS-state on the other hand both $e_g$ levels and two of the three $t_{2g}$ levels are occupied; again $\nu = 3$. (The five-fold $S_z$ degeneracy leading to $\nu = 15$ in Ref. [6] is lifted by the field.)

For a field of 5.8 T $\delta = 7.80$ K. Putting this value of $\delta$ and $m(100\text{ K}) = 0.09(1)\,\mu_B$ (8) in Eq. (9) one obtains $\Delta_I = 32^{+115}_{-32}$ and $\Delta_H = 430^{+34}_{-20}$, at $T = 100$ K. The superscripts and subscripts give the estimated standard deviations. In spite of very high uncertainty in $\Delta_I$, its value ishows clearly that it is the IS state which is is excited at $T \approx 100$ K. Actually $\Delta_H$ determined from $m(285\text{ K})$ is even higher; we estimate it as $\Delta_H \approx 1000$ K, a value obtained without correction for the IS contribution which is difficult to calculate because of the large uncertainty in $\Delta_I$.

The covalent moment on the ligand $O^{2-}$ at 100 K for the models 3 and 4 (table 2) is only about 3 times greater than its standard deviation. This is not sufficient to allow any conclusion to be drawn about its temperature dependence such as that invoked [20] to account for the negative cooperative effect preventing long-range spin ordering in the excited state. The values obtained are not inconsistent with the results [25] from which a covalent moment amounting to a few percent of the $Co^{3+}$ moment can be estimated. The statistical precision would need to be improved by a factor of about 5 to allow any temperature variation to be verified. Such an improvement might be possible with a larger crystal.

In summary, we have shown that the spin-state transition at $T \approx 100$ K is related to the LS $\rightarrow$ IS one. There is some evidence for covalent spin density on the ligands amounting to about 10% of the cobalt moment, although it is close to the limit of significance. An experiment using a larger crystal and longer counting times should enable the temperature dependence of the covalent contribution to be determined.

**Acknowledgments**

This work was carried out in framework of the INTAS project (Grant N 01-0278). We are deeply indebted to INTAS for the Grant, which has made possible this study. We are grateful to the project coordinator Prof. A. Furrer for the fruitful scientific discussions and excellent management. Partial supports from the Russian Foundations for Basic Researches (Project N° 05-02-17466-a), by the grant SS-1671.2003.2 and by the NATO grant PST CLG 979369 are acknowledged. Two of us, V.P. Plakhty and S.V. Gavrilov, are indebted to ILL for hospitality.